\documentclass[doublecol]{epl2} 

\title{Collective chaos in pulse-coupled neural networks}
\shorttitle{Collective chaos} 

\author{Simona Olmi\inst{1,2,3} \and Antonio Politi\inst{1,3} \and
 Alessandro Torcini\inst{1,2,3}}
\shortauthor{S. Olmi \etal}

\institute{                    
  \inst{1} Istituto dei Sistemi Complessi, CNR, via Madonna del Piano 10,
   I-50019 Sesto Fiorentino, Italy\\
  \inst{2} INFN Sez. Firenze, via Sansone, 1 - I-50019 Sesto Fiorentino, Italy\\
  \inst{3} Centro Interdipartimentale per lo Studio delle Dinamiche Complesse,
   via Sansone, 1 - I-50019 Sesto Fiorentino, Italy\\
}


\abstract{
We study the dynamics of two symmetrically coupled populations of
\revision{identical}
leaky integrate-and-fire neurons characterized by an excitatory coupling.
Upon varying the coupling strength, we find symmetry-breaking transitions that
lead to the onset of various chimera states as well as to a new regime, where
the two populations are characterized by a different degree of synchronization.
Symmetric collective states of increasing dynamical complexity are also
observed. The computation of the the finite-amplitude Lyapunov exponent
allows us to establish the chaoticity of the (collective) dynamics in a
finite region of the phase plane. The further numerical study of the standard
Lyapunov spectrum reveals the presence of several positive exponents, indicating
that the microscopic dynamics is high-dimensional.}

\begin{document}

\maketitle

\section{Introduction}

Understanding the collective motion of ensembles/networks of oscillators is 
crucial in many contexts, starting from neuronal circuits~\cite{buszaki}. So
far, most of the efforts have been devoted to the characterization of strong
forms of synchronization. However, more subtle phenomena, like the onset of
collective motion in an ensemble of (chaotic) units, which behave in a seemingly
uncorrelated way can also play a relevant role for information encoding.
Collective chaos, meant as irregular dynamics of coarse-grained observables,
has been found in ensembles of fully coupled one-dimensional
maps~\cite{kane_coll,cencio_coll} as well as in two-dimensional continuous-time
oscillators~\cite{matthews,kura_coll,hakim_coll}. In both classes of models, the
single
dynamical unit can behave chaotically under the action of a periodic forcing
(in non-invertible maps, there is even no need of a periodic forcing). 
What does it happen 
in ensembles of phase-oscillators which cannot become chaotic under
the action of any forcing? The evolution of a (formally infinite) population of
oscillators is ruled by a self-consistent (nonlinear) functional equation for
the probability density. Given the infinite dimensionality of the model, the
population could, in principle, behave chaotically, irrespective of the
``structure" of the single oscillators. In spite of this potentiality, 
\revision{only a few examples of low-dimensional}
chaotic collective motion \revision{have} been found in ensembles of
phase-oscillators \revision{~\cite{chaos,watanabe}}. One reason is that 
most of the models so far investigated
are based on sinusoidal force fields (in the following we refer to them as
to sinusoidal oscillators); in this setup, there is little space for
a high dimensional dynamics, since no matter how many oscillators are involved,
there are always $N-3$ constants of motion \cite{watanabe}. 
This ``degeneracy" is
not present in typical pulse-coupled networks of neurons, where different force
fields are usually assumed. A prototypical example is that of leaky
integrate-and-fire (LIF) neurons, characterized by a linear force field.
It is, in fact, not suprising that the first instance of a nontrivial
collective motion has been found in an ensemble of LIF neurons. We refer to
Partial Synchronization (PS) \cite{vvres}, a regime characterized by a periodic
macroscopic dynamics and a quasiperiodic microscopic motion, with the additional
subtlety that the average inter-spike interval of the single neurons differs
from the period of the collective variable. More recently, this type of
behaviour has been observed also in a population of sinusoidal oscillators,
in the presence of a suitable nonlinear coupling\cite{piko}.

The only and quite striking evidence of an irregular collective dynamics has
been recently found in an ensemble of LIF neurons in the presence of a random
distribution of the input currents~\cite{luccioli} (this setup, where the single
neurons are characterized by different spiking rates, is analogous to that of
the usual Kuramoto model, where the single oscillators have different bare
frequencies). On the other hand, the model studied in \cite{luccioli} has the
further peculiarity of exhibiting a negative maximal Lyapunov exponent --
it is, in fact, an example of {\it stable chaos}~\cite{poltor}. However, in the
absence of disorder, no example of irregular dynamics has yet been found.

A slightly more complex but meaningful setup is that of two symmetrically
coupled populations of identical oscillators. This is the simplest instance of
``network-of-networks" that is often invoked as a paradigm for neural systems
\revision{~\cite{netnet}}.
With reference to sinusoidal oscillators, this setup has revealed the
onset of chimera states (one of the two populations is fully synchronized,
while the oscillators of the other one are not synchronized at all\cite{chimera}),
as well as more complex macroscopic states with periodic \cite{abrams} and
quasi-periodic \cite{pikov_chimera} collective oscillations.
In the present Letter we study the two-population setup with reference to LIF
neurons for different values of the coupling strengths between and within the
two populations. We find various kinds of symmetry broken states some of which
are similar to those observed in \cite{abrams,pikov_chimera} and a new
one, where the two populations are both partially synchronized, but with a
different degree. More interesting is the parameter region where the
collective motion is chaotic, as indicated by the finite-amplitude Lyapunov
exponent (FALE) \cite{fsle} and confirmed by the computation of the standard
Lyapunov spectrum which reveals the existence of several positive exponents.

\section{The model}

We consider two fully coupled networks, each made of $N$ LIF oscillators.
Following Refs.~\cite{zillmer2}, the membrane potential $x_j^{(k)}(t)$
of the $j-th$ oscillator ($j=1,\ldots,N$) of the $k$th population ($k=0,1$)
evolves according to the differential equation,
\begin{equation}\label{eq:single}
\dot{x}_{j}^{(k)}(t)= a-x_{j}^{(k)}(t)+g_s E^{(k)}(t) +g_c E^{(1-k)}(t)
\end{equation}
where $a >1$ is the suprathreshold input current, while $g_s > 0$  and $g_c > 0$
gauge the self- and, resp., cross-coupling strength of the excitatory
interaction. \revision{Whenever the membrane potential
reaches the threshold $x_j^{(k)}=1$, it is reset to 
$x_j^{(k)}=0$, while a so-called $\alpha$-pulse is sent and instantaneously
received by all the neurons. The field $E^{(k)}(t)$ represents the linear
superposition of the pulses emitted within the population $k$ in the past.
It can be shown \cite{zillmer2} that $E^{(k)}(t)$} satisfies the differential
equation
\begin{equation}
\label{eq:E}
  \ddot E^{(k)}(t) +2\alpha\dot E^{(k)}(t)+\alpha^2 E^{(k)}(t)=
  \frac{\alpha^2}{N}\sum_{j,n} \delta(t-t_{j,n}^{(k)}) \ ,
\end{equation}
where $t_{j,n}^{(k)}$ is the $n$th spiking time of the $j$th neuron within the
population $k$, and the sum is restricted to times smaller than $t$.
In the limit case $g_s=g_c=g$, the two populations can be seen as a single
one made of $2N$ neurons with an effective coupling constant $G= 2g$.

The degree of synchronization can be quantified by introducing the typical
order parameter used for phase oscillators
$r^{(k)}(t) =\left| \langle \exp[i \theta_j^{(k)}(t)]\rangle \right|$, where 
$\theta_j^{(k)}$ is the phase of the $j$th oscillator, that can be properly
defined \revision{by suitably rescaling the time variable~\cite{winfree},
$\theta_j^{(k)}(t) = 2\pi (t-t_{j,n}^{(k)})/(t_{q,n}^{(k)}-t_{q,n-1}^{(k)})$, 
where $n$ identifies the last spike emitted by the $j$th neuron, while
$q$ indicates the neuron that has emitted the last spike.} One can
verify that this phase is bounded between 0 and $2\pi$, as it should.
It is interesting to see that the application of this definition to the PS
regime described by van Vreeswijk \cite{vvres} reveals that the order parameter
fluctuates periodically. In other words PS differs from the regime observed in
the Kuramoto model above the synchronization threshold, where the order
parameter is constant in time \cite{kurabook}.

\section{Phase Diagram}

\begin{figure}
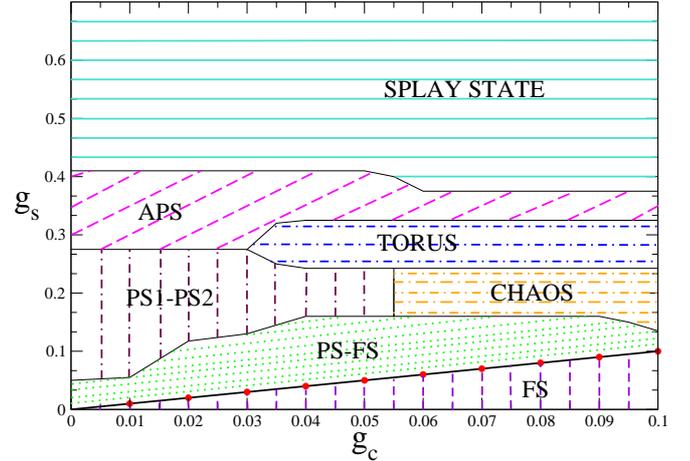

\onefigure[draft=false,clip=true,height=0.34\textwidth]{fig1.eps}
\caption{(Color Online) 
\revision{Phase diagram in the $(g_c,g_s)$-plane of the model
(\ref{eq:single},\ref{eq:E}) for $a=1.3$ and $\alpha=9$. FS indicates Full
Synchronization (both populations fire at the same
time); PS-FS indicates that the two populations are is the FS and PS regimes,
respectively; PS1-PS2 indicates that both populations are in a PS regime,
although with a different degree of synchronization; APS indicates Antiphase
Partial Synchronization, i.e. the two fields exhibit the same behaviour though
being in antiphase; TORUS indicates a collective quasi-periodic motion; finally,
CHAOS indicates collective chaotic motion.}}
\label{fig.1}
\end{figure}

The equations have been integrated by extending the event-driven approach
described e.g. in \cite{zillmer2}. In practice, the (linear) equations of motion
are solved analytically in between two consecutive spike-emissions, obtaining a
suitable map. Since the ordering of the single-neuron potentials does not
change within each population, the next firing event can be easily
determined by comparing the neurons that are closest to threshold within each of
two populations. In spite of the conceptual simplicity and the effectiveness
of the code, one must be nevertheless careful in handling nearly singular cases,
when many neurons almost cluster together. In order to avoid the spurious
clustering, due to numerical roundoff, we have changed variables, introducing
and monitoring the logarithm of the difference of the membrane potentials of
two successive neurons. This requires some care in defining the right number of
variables: given $N$ potentials, one naturally has $N-1$ differences that have
to be complemented by a proper $N$th variable (for more details
see \cite{olmi_lungo}).

\begin{figure}
\onefigure[draft=false,clip=true,height=0.34\textwidth]{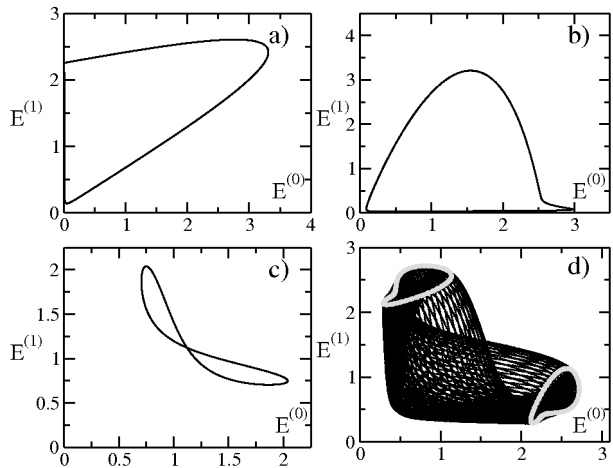}
\caption{The macroscopic attractors displayed by
reporting the fields $E^{(0)}$ vs $E^{(1)}$ for four different non chaotic
phases, namely (a) PS-FS $(g_c=0.07$, $g_s=0.1)$, 
(b) PS1-PS2 $(g_c=0.02$, $g_s=0.17)$, (c) APS $(g_c=0.07$, $g_s=0.35)$ 
and (d) TORUS $(g_c=0.07$, $g_s=0.3)$ (for the 
exact definitions see the text). The \revision{grey} curves reported in the panel (d)
are the Poincar\'e sections obtained by imposing that the sum 
$E^{(0)}+E^{(1)}$ is maximal.
}
\label{fig.1bis}
\end{figure}

\begin{figure}
\onefigure[draft=false,clip=true,height=0.34\textwidth]{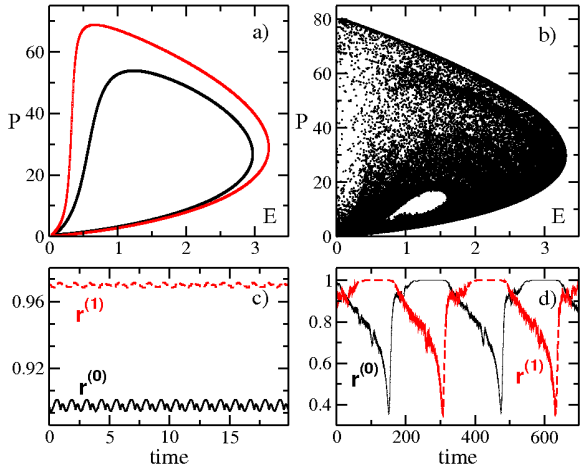}
\caption{(Color Online) Macroscopic attractors displayed by reporting
$P \equiv E+\alpha \dot E$ vs $E$ for a PS1-PS2 state (a) and 
a chaotic phase (b), the time evolution of the corresponding order 
parameters $r^{(0)}$ and $r^{(1)}$ is also reported in (c) and (d). 
The variables corresponding to population 0 (resp. 1) 
are shown as \revision{black solid lines} (resp. \revision{red dashed lines})
\revision{in (c) and (d); while in (a) the internal black (resp. external red) curve
refers to population 0 (resp. 1) and in (b) an unique attractor has been 
reported for clarity reasons, since the two attractors are overlapping}.
The data reported in (a) and (c) refer to $g_c=0.02$ and $g_s=0.17$,
while those shown in (b),(d) to $g_c=0.08$ and $g_s=0.16$.
}
\label{fig.2}
\end{figure}

The phase plane $(g_c,g_s)$ shown in Fig. ~\ref{fig.1} has been obtained by
studying the model (\ref{eq:single},\ref{eq:E}) for $a=1.3$ and $\alpha=9$.
The diagram is semiquantitative in the sense that a much more detailed work
would be needed to identify exactly the stability borders of the different
regimes.
Along the diagonal ($g=g_s=g_c$) the model reduces to that for a single
population with coupling strength $G=2g$. For our choice of $a$ and $\alpha$
values, the system exhibits PS, since we are below the critical value
$G_0=0.425$ \cite{zillmer2} above which the splay state is stable (the splay
state is a regime characterized by a constant spiking rate and thereby a
constant field, i.e. no collective dynamics). Below the diagonal, the
evolution is still symmetric but fully synchronized (FS),
i.e. all neurons of both populations fire together. More intriguing is the
region above the diagonal, that is characterized by a spontaneous symmetry
breaking: one population fully synchronizes, while the other is in a PS regime,
i.e. we are in presence of a generalized chimera state (here termed PS-FS).
This can be appreciated by looking at the synchronization parameter $r^{(k)}$
of the two populations, one of which is equal to one, while the other oscillates
periodically close to 0.8. By following \cite{abrams}, this state can be
classified as a {\it periodically breathing chimera}. In this regime, the two
populations are characterized by a microscopically periodic and quasi-periodic
behaviour, respectively. In spite of this qualitative difference, the two
(macroscopic) fields $E^{(0)}$ and $E^{(1)}$ are both periodic and phase locked
(see Fig.~\ref{fig.1bis}a). This means that the neurons subject to two different
linear combinations of $E^{(0)}$ and $E^{(1)}$ behave differently: a population
locks with the forcing field, while the other one behaves quasi-periodically.
Another even more interesting symmetry broken state can be observed for larger
$g_s$-values and $ g_c < 0.055$; in this case both populations exhibit PS, 
but their dynamics take place over two different attractors with two different
degrees of synchronization (PS1-PS2 regime), as shown in Fig.~\ref{fig.2}a. Like in the PS-FS
regime, the two fields behave periodically (with the same period) and are phase
locked, as it can be appreciated by looking at the closed curve $E^{(0)}$ versus
$E^{(1)}$ in Fig. \ref{fig.1bis}b.
However, at variance with PS-FS, here both populations exhibit quasi-periodic
motions.
In other words we are in presence of a different symmetry breaking, where two
populations with distinct quasi-periodic motions spontaneously emerge. 
For yet larger $g_s$ values, the equivalence between the collective dynamics of
the two population is restored, the only difference being a phase shift between
the two fields, which oscillate in antiphase and this is why we term this regime
{\it Antiphase Partial Synchronization} (APS). In the APS phase, for finite $N$,
the \revision{instantaneous} maximum Lyapunov exponent strongly fluctuates and we cannot
rule out the possible existence of some form of weak chaos, analogous to the
one discussed in~\cite{simo} for a model of diluted neural network, i.e. a
chaotic behaviour that disappears in the thermodynamic limit.
In a strip above the chaotic region (discussed below), one can observe
collective quasiperiodic motion. This means that the quasiperiodic motion of the
fields is accompanied by a dynamics of the single neurons along a torus $T^3$.
An analogous regime has been previously reported in~\cite{kura_coll} in the
context of a population of coupled Stuart-Landau oscillators. Here, we find it
in a model where the single units are described by a single variable.
Furthermore, we have characterized the motion on the macroscopic $T^2$ attractor
reported in Fig. \ref{fig.1bis}, by estimating the winding numbers for various
values of the coupling strengths. We find that the winding number is
typically quite small -- on the order of $\sim 10^{-2}$ -- but independent of
the system size, indicating that the torus survives in the thermodynamic limit.
Finally, for yet larger $g_s$-values both populations converge towards a splay
state. This is not surprising, as we already know that for the chosen 
$\alpha$- and $a$-values, the splay state is stable in a single population of
neurons for $G > G_0 \equiv 0.425$. 

\begin{figure}
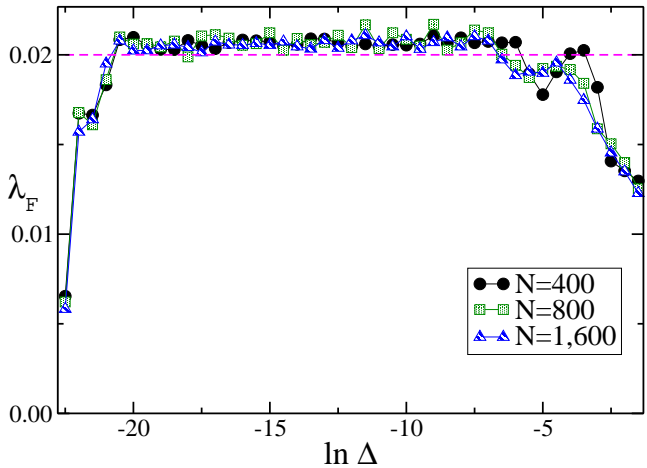

\onefigure[draft=false,clip=true,height=0.34\textwidth]{fig4.eps}
\caption{ 
Finite amplitude Lyapunov exponents $\lambda_F$ versus the logarithm of the
perturbation amplitude $\Delta$ for three different system sizes: namely,
$N=400$ (black circles), $N=800$ (green squares) and $N=1,600$ (blue triangles).
The amplitudes $\Delta$ have been estimated by considering the euclidean distance
among the perturbed and unperturbed fields. The dashed (magenta) line indicates
the maximal microscopic Lyapunov exponent $\lambda_1$ for $N=1,600$ obtained by
following the orbit over a time span containing $10^8$ spikes and after discarding a 
transient composed by $10^6$ spikes. $\lambda_F$ has been estimated by 
averaging over $25,000 - 50,000$ different trajectories. The results have been obtained
for $g_c=0.08$ and $g_s=0.16$.
}
\label{fig.3}
\end{figure}

\section{Collective Chaos}

In a limited region above the diagonal and for $g_c > 0.055$ the collective
behaviour is irregular, this is clearly seen by observing the two macroscopic
attractors, corresponding to the two populations, reported in Fig. \ref{fig.2}b.
Nonetheless, from Fig. \ref{fig.2}c it appears that the associated order parameters behave 
almost periodically, but this periodicity is only apparent since it
reflects only the larger time scale present in the system, the one associated to 
the modulation of the fields $E^{(k)}$, while the irregularity of the dynamics
are more evident at smaller time scales. In particular, \revision{at least two other} 
time scales 
are present: a scale ${\cal O}(1)$ associated to the firing period of
each specific neuron and a scale ${\cal O}(1/N)$ corresponding to the 
interspike interval between two successive spike emissions in the network.
In order to make a quantitative assessment of the chaoticity we have first studied the FALE $\lambda_F$. The
FALE can be determined from the growth rate of a small finite perturbation
for different amplitudes $\Delta$ of the perturbation itself (after averaging
over different trajectories)~\cite{fsle}. This is done by randomly perturbing
the coordinates (both fields and the membrane potentials of the two populations)
of a generic configuaration on the attractor. The results for $g_s=0.16$,
$g_c=0.08$ and three different system sizes are plotted
in Fig.~\ref{fig.3}. For small $\Delta$ values, $\lambda_F$ grows with $\Delta$,
since the perturbation needs first to converge towards the most expanding
direction, while the final drop is a manifestation of the saturation
of the perturbation amplitude. It is the height of the intermediate plateau
which measures the amplitude of the FALE. Since the height is independent of
$N$, one can conjecture that the collective motion is chaotic and stays chaotic
in the thermodynamic limit. Besides the data reported in Fig.~\ref{fig.3},
we have checked that the plateau height is not influenced by changing the
amplitude of the initial perturbation and by performing tests up to $N$=6,400.

It is instructive to compare $\lambda_F$ with the maximum $\lambda_1$ of the
standard Lyapunov spectrum. For small $\Delta$ values, the two indicators should
coincide, but there is no reason for the agreement to persist at larger
amplitudes. In fact, a second plateau has been detected in the context of
globally coupled maps\cite{kane_coll, cencio_coll}.  The plateau
occurring for larger $\Delta$'s has been interpreted as an indication that
the chaoticity of collective variables differs from that of the microscopic
ones. Since, in the present case, we observe only a single plateau, it is
crucial to verify its compatibility
with $\lambda_1$. The horizontal dashed line in Fig.~\ref{fig.3} corresponds
to the Lyapunov exponent determined for $N$=1,600 (for the dependence of
$\lambda_1$ on $N$, see below). The two indicators are consistent with each
other and this means that collective variables are as chaotic as the
microscopic ones.

Having established the existence of one unstable direction, the next question
is to determine how many such directions are present. Unfortunately, the
concept of FALE allows to determine just one exponent. As a consequence, we
turn our attention to the usual Lyapunov spectrum, well aware that the
``microscopic" exponents do not necessarily reproduce the chaoticity of the
``macroscopic" variables.

\begin{figure}
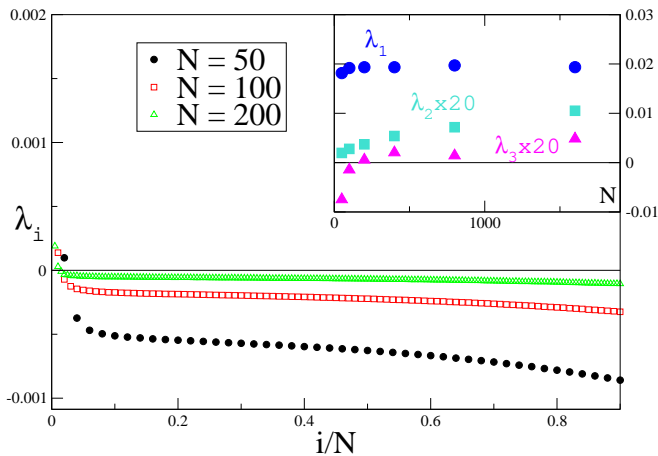

\onefigure[draft=false,clip=true,height=0.33\textwidth]{fig5.eps}
\caption{(Color Online) Lyapunov spectra $\lambda_i$ versus $i/N$ 
for three $N$-values: namely, $N=50$ (filled black circles), $N=100$
(empty red squares) and $N=200$ (empty green triangles). In the inset
the first three Lyapunov exponents are reported as a function of $N$: 
$\lambda_1$ (blue circles),
$\lambda_2$ (turquoise squares) and $\lambda_3$ (magenta triangles).
The Lyapunov esponents have been obtained by following the dynamics in the
real and tangent space for a time span containing $10^8 - 5\times 10^9$
spikes, after discarding a transient period of $10^6$ spikes.
The reported results refer to $g_c=0.08$ and $g_s=0.16$.
}
\label{fig.4}
\end{figure}

In the present model, the Lyapunov spectrum $\{\lambda_i\}$ is composed of
$N+3$ exponents ($i=1,\dots, N+3$). The phase space dimension is in fact
equal to $N+4$ ($N$ potentials plus 2 two-dimensional equations for the fields),
but the direction corresponding to the zero-Lyapunov exponent is implicitly
eliminated as a consequence of taking the Poincar\'e section \cite{zillmer2}. 
In Fig.~\ref{fig.4} we have plotted the first part of the spectrum (with
the exception of $\lambda_1$, for the clarity of presentation) with the usual
normalization $i/N$ of the $x$ axis. The figure clearly indicates that the
spectrum becomes increasingly flat and converges to zero. This can be
understood in the following way. In the thermodynamic limit, the dynamics of
globally coupled identical oscillators can be viewed as that of single
oscillators forced by the same (self-consistent) field \cite{kura_coll}. As a
result, in a first approximation, we expect all Lyapunov exponents to be equal
to the conditional Lypunov exponent \revision{$\lambda_c$ obtained by forcing a
single LIF neuron (identical to the others) with the self-consistent fields
(obtained by integrating the whole ensemble). We have found that $\lambda_c=0$
(within numerical accuracy).} This justifies why the increasingly flat Lyapunov
spectrum converges towards zero. Moreover, since a LIF neuron is described by a
single variable, \revision{under no circumstance, $\lambda_c$ can be strictly}
positive. On the other hand, $\lambda_c$ can be negative, and we expect this to
occur whenever the given population exhibits full
synchronization. Since we know that no synchronization is observed in the
chaotic regime, we can conclude that $\lambda_c$ is not just very small, but
it must be exactly equal to zero.

Having established that the Lyapunov spectrum converges to zero, it is interesting
to investigate its scaling behaviour. By comparing the spectra obtained for
different system sizes, it turns out that they scale as $1/N^\beta$, although
we are unable to extrapolate the value of $\beta$ from the analysis of the
relatively small systems that we have simulated. By comparing the spectra
obtained for $N=100$ and $200$ we can at most guess that $\beta\approx 1.5$. This
value is not far from $\beta=2$, found analytically while studying the
splay-state stability in single populations of LIF neurons~\cite{calamai} and
numerically for the PS state~\cite{simo}.

Finally, we turn our attention to the first part of the spectrum, where the
flatness hypothesis does not hold. More precisely, we investigate 
the $N$-dependence of the first three Lyapunov exponents which can be computed
for larger lattices (up to $N$=1,600). The maximal Lyapunov exponent appears to
converge to a finite asymptotic value $\lambda_1 = 0.0195(3)$. On the other
hand, the second and third exponents grow systematically with $N$, both becoming
positive for $N > 200$, with no clear evidence of an eventual saturation.
These results suggest that the microscopic chaos is high-dimensional
(there is no reason to believe that the number of positive exponents is just
equal to three). However, we cannot tell whether the number of positive
exponents is extensive (proportional to $N$) or sub-extensive.

\section{Conclusions}

We have studied two symmetrically coupled populations of leaky
integrate-and-fire neurons for different values of the self- ($g_s$)
and cross-($g_c$) coupling-strength. Some of the collective phenomena that
we have identified are quite similar to those observed in the two-population
setup of Kuramoto-like oscillators\cite{note}. This is not surprising, since it
is known that an ensemble of LIF neurons is equivalent, in the weak coupling
limit, to the Kuramoto model \cite{hansel_1995}, the only difference being that
the coupling function is not purely sinusoidal. The onset of PS in both classes
of models suggests that the equivalence can be extended to larger coupling
strengths. However, since PS can be obtained in the Kuramoto setup only by
invoking a more general kind of coupling \cite{piko}, it is legitimate to
conclude that the relationship \revision{is more complicated} than that suggested by the
study of the weak-coupling limit. In fact, in this Letter we have found
new dynamical regimes, such as a different PS dynamics for the two populations.
A yet more intriguing phenomenon is the collective chaos that we have
recognized as such from the computation of the finite-amplitude Lyapunov
exponent.
Altogether, a general question still stands: \revision{To what extent are pulse-coupled
oscillators} equivalent to Kuramoto-like models? The identification of the
mutual relationship would be highly beneficial in both areas.

Another still open question is that of the degree of chaoticity of the
collective dynamics. This problem is also connected to that
of the asymptotic structure of the Lyapunov spectrum in the thermodynamic limit.
A rough argument suggests that the spectrum should be flat and this is indeed
approximately seen in the numerics. However, the evident deviations observed in
the vicinity of the maximum strongly suggest that the argument need be refined.
Altogether, we observe that the Lyapunov spectrum includes the FALE. This means
that the evolution of not-so-small perturbations does not add anything, to that
of infinitesimal ones and implies that the standard Lyapunov analysis is rich
enough to account for the collective behaviour as well. This was not a priori
obvious. The computation of the FALE $\lambda_F$ in globally coupled
maps\cite{cencio_coll}
reveals a different scenario, where $\lambda_F$ differs from the maximum
Lyapunov exponent. On the other hand, the more recent study of globally coupled
Stuart-Landau oscillators \cite{ginelli} provides an example where, like here,
the chaoticity of the collective motion can be inferred from the Lyapunov
spectrum.
Moreover, what can we say about the role of the second and third exponents that
are found to be positive as well? Do they contribute to the microscopic
dynamics only, or also to the macroscopic one? A careful analysis based on the
study of the corresponding covariant Lyapunov vectors \cite{lyvector} might help
to clarify this point. An alternative and more direct approach could be that of
(numerically) integrating the self-consistent dynamical equation for the
probability densities of the membrane potentials in the two families. However,
it is not easy to pursue this latter perspective: because of the occasional
formation of strongly clusterized states, it is necessary to partition the
phase space into a huge number of small cells.

\acknowledgments
We thank M. Cencini, F. Ginelli, A. Pikovsky \revision{and S.H. Strogatz}
for a careful reading of the manuscript.
This research project is part of the activity
of the Joint Italian-Israeli Laboratory on Neuroscience
funded by the Italian Ministry of Foreign Affairs and it
has been partially realized thanks to the support of CINECA
through the Italian Super Computing Resource Allocation
(ISCRA) programme\revision{, project ECOSFNN}.


\begin{thebibliography}{0}
\bibitem{note}
By Kuramoto-like we mean sinusoidal oscillators with some sort
of continuous-time coupling that depends on the phase-differences.
\end{thebibliography}
\end{document}